\documentclass[journal]{IEEEtran}
\usepackage[utf8]{inputenc}
\usepackage[T1]{fontenc}
\usepackage{units}
\usepackage{amsmath}
\usepackage{amsthm}
\usepackage{graphicx}
\usepackage[english]{babel}
\usepackage[bookmarks=true,bookmarksnumbered=true,bookmarksopen=true,bookmarksopenlevel=1,
 breaklinks=false,pdfborder={0 0 0},pdfborderstyle={},backref=false,colorlinks=false]
 {hyperref}
\hypersetup{pdftitle={Your Title},
 pdfauthor={Your Name},
 pdfpagelayout=OneColumn, pdfnewwindow=true, pdfstartview=XYZ, plainpages=false}

\makeatletter

\providecommand{\tabularnewline}{\\}

\let\oldforeign@language\foreign@language
\DeclareRobustCommand{\foreign@language}[1]{%
  \lowercase{\oldforeign@language{#1}}}
\theoremstyle{plain}
\newtheorem{thm}{\protect\theoremname}

\usepackage[caption=false,font=footnotesize]{subfig}

\ifdefined\showcaptionsetup
 \PassOptionsToPackage{caption=false}{subfig}
\fi
\usepackage{subfig}
\makeatother

\providecommand{\theoremname}{Theorem}
\providecommand{\corollaryname}{Corollary}
\renewcommand{\qedsymbol}{\hfill \rule{1.5ex}{1.5ex}}

\begin{document}
\title{Theoretical Bounds for Optimized Doppler-Based Motion Detection in UHF-RFID
Readers}
\author{Clemens~Korn,~and~Joerg~Robert,~\IEEEmembership{Member,~IEEE}\thanks{Clemens~Korn is with the Dependable M2M Research Group, Technische Universitaet Ilmenau, 98693 Ilmenau, Germany, and with the Self-Powered Radio Systems Department, Fraunhofer IIS, 90411 Nürnberg, Germany,
e-mail: \protect\href{mailto:clemens.korn@iis.fraunhofer.de}{clemens.korn@iis.fraunhofer.de}}\thanks{Joerg~Robert is with the Friedrich-Alexander Universität Erlangen-Nürnberg (FAU), Information Technology (Communication Electronics), 91058 Erlangen, Germany, and with the Fraunhofer IIS, 90411 Nürnberg, Germany, e-mail:
\protect\href{mailto:joerg.robert@ieee.org}{joerg.robert@ieee.org}}}
\markboth{}{Clemens Korn \MakeLowercase{\emph{et al.}}: Theoretical Bounds for Optimized Doppler-Based Motion Detection in UHF-RFID
Readers}

\IEEEaftertitletext{This work has been submitted to the IEEE for possible publication. Copyright may be transferred without notice, after which this version may no longer be accessible.}

\maketitle

\global\long\def\sgn{\mathrm{{sgn}}}
\global\long\def\erf{\mathrm{{erf}}}
\global\long\def\erfinv{\mathrm{{erfinv}}}
\global\long\def\argmax{\mathrm{{argmax}}}
\global\long\def\rect{\mathrm{{rect}}}

\begin{abstract}
Radio Frequency Identification (RFID) is a widely used technology for identifying and locating objects equipped with low-cost RFID transponders (tags).  
UHF (Ultra High Frequency) RFID operates in frequency bands around 900\,MHz and supports communication distances of up to 15\,m between the reader and the tag.
Reliable motion detection is therefore a highly relevant feature in modern logistics -- for example, to determine whether a tag is actually placed on a conveyor belt or merely in its vicinity.

A promising approach for accurate motion detection is the use of the Doppler effect.  
Some state-of-the-art UHF-RFID readers already support Doppler shift measurements.
However, their measurement accuracy is insufficient for many applications.  
In this paper, we propose an optimized method for the precise Doppler shift estimation using existing RFID systems -- an essential step toward enabling RFID-based motion detection in future logistics.  
Further, we also derive the theoretical bounds for Doppler-based motion detection in UHF-RFID systems based on the Cramer-Rao Lower Bound.  
These bounds analyze the influence of tag signal strength, signal duration, and the intervals between multiple tag replies on the performance of motion detection and speed estimation algorithms.  
In addition, we establish theoretical limits that account for hardware constraints in current UHF-RFID readers.

The results of this work provide valuable insights into the limitations of Doppler-based motion detection and support system-level performance optimization.  
They enable prediction of achievable performance based on reader noise figure, aiding in the design and tuning of RFID systems.

\end{abstract}


\IEEEpeerreviewmaketitle{}

\section{Introduction}
\IEEEPARstart{R}{adio} Frequency Identification (RFID) is a technology used to identify and locate objects equipped with low-cost RFID transponders (tags) \cite{finkenzeller2015rfid}.
UHF-RFID (Ultra-High Frequency RFID) systems operate in frequency bands around 900\,MHz over distances up to 15\,m between the reader and the tags.
This technology is widely adopted in industry and logistics, allowing the identification, localization, and tracking of goods across warehouses and production lines, with billions of tags deployed worldwide.


Despite its advantages, UHF-RFID still faces unresolved challenges in applications that require precise tag positioning or movement tracking, beyond simply detecting its presence within the reader's range. 
A typical example is a warehouse entrance gate equipped with an RFID reader to monitor goods moving in and out.
In such scenarios, achieving reliable detection can be difficult, as the reader may inadvertently detect nearby tags that are not actually going through the gate due to its wide reading range. Reducing the reading range is often impractical. 
To address this, various solutions have been proposed to enhance tag localization within the reader's range.
These include techniques such as ranging with multiple reader antennas or tracking the tag's phase on multiple read-outs \cite{NikitinPhaseBasedSpatialIdentification,ScherhaeuflPhaseOfArrival,BuffiAPhaseBased,DiGiampaoloRangeAndBearing,ZhangVehicularLocalization,FanMobileFeatureEnhanced,VossiekUHFRFIDLocalization}.

A cheaper approach that can tackle these and other problems without
additional hardware -- like multiple antennas or additional readers --
is to implement a motion detection that is based on the Doppler effect~\cite{molisch2012wireless}.
This allows to estimate the current speed of the tag relative to the
reader antenna by means of the time variance of the received tag signals and digital signal processing.
The big advantage of the Doppler approach is that it is a solution purely based on digital signal processing. 
Hence, it does not require any additional hardware -- such as additional antennas or multiple readers -- and is therefore comparably cheap.
Generally, RFID systems are perfectly suited for Doppler estimation with relatively small Doppler frequency offsets, as the transmission and reception take place in the same physical device.
Consequently, frequency offsets or effects due to phase noise between the transmitter and the receiver -- which would impact the estimation precision -- do practically not exist.

There are already RFID readers on the market that
provide Doppler shift measurements~\cite{Xu2018}, an example is the company Impinj\footnote{Impinj Revolution Low Level User Data Support, 2013, revision 3.0.,
available: https://support.impinj.com/hc/en-us/article\_attachments/200774268/SR\_
AN\_IPJ\_Speedway\_Rev\_Low\_Level\_Data\_Support\_20130911.pdf (accessed: March 12, 2025)}.
Unfortunately, these Doppler measurements are currently not precise enough for the majority of applications.
The reasons are the weak reception levels and the short transmit durations of the tags.
Nevertheless, there exists potential for significantly improving the precision of Doppler measurements.
A promising approach is to optimize the communication between the reader and the tag in a way that optimizes the estimation of the Doppler shift.
A simple approach is presented in~\cite{Zhai2018}, where the authors use multiple reads of a tag to improve the estimation precision.

The suitability of Doppler estimation for the aforementioned use cases has already been demonstrated in the literature. In~\cite{Zhai2018}, the phase of RFID tag replies -- i.e. the Doppler shift -- was used to detect the speed of moving tags mounted on vehicles. 
Similarly, the Doppler effect was used to estimate the speed of passing cars in~\cite{Yang2019}, and to track RFID labels mounted on a toy car and a paper box on a conveyor belt in~\cite{Tesch2015}. 
The conveyor belt scenario was further explored in~\cite{Xu2018}, which proposed a Doppler-based classification algorithm for tags, incorporating the strength of the received signal for motion detection. 
In contrast, \cite{Yu2011} investigated a setup in which the reader, as the moving component, detected the motion of the tag by estimating the Doppler shift. Furthermore, \cite{Ahmad2020} introduced a novel tag design enabling Doppler shift measurements between different RFID tags.
These studies demonstrate the feasibility of detecting tag motion and highlight the results achieved with their specific approaches.
However, to the best of our knowledge, no prior work has systematically optimized the performance of Doppler estimation in UHF-RFID or derived bounds on its achievable performance.

In this paper, we present an in-depth analysis of the required Doppler estimation precision for reliable motion detection. 
To achieve this, we use the Modified Cramér-Rao Bound (MCRB)~\cite{MCRBAndrea}, which estimates the achievable variance of an unbiased estimator. 
The objective of this work is to establish a theoretical foundation for the development of new approaches to optimize UHF-RFID systems in terms of Doppler shift estimation precision and Doppler-based motion detection reliability. 
In addition, it provides a practical framework for evaluating the feasibility of specific use cases.

The paper is organized as follows. Section \ref{sec:System_Model} introduces the system model and outlines the assumptions used for the subsequent theoretical analyses.
Section \ref{sec:Minimum_Required_Estimation_Precision} examines the precision of Doppler shift estimation required to reliably distinguish between a static tag and a moving tag. 
In Section \ref{sec:MCRBSection}, the maximally achievable estimation precision is analyzed using the Modified Cramér-Rao Bound.
Section \ref{sec:Boundaries-of-the-Motion-Detection} combines the findings of the previous sections to determine the tag speeds and tag signal strengths necessary to reliably differentiate between static and moving tags. 
Finally, Section \ref{sec:Conclusions} concludes the paper and offers perspectives for future research.
The extensive annex includes detailed derivations.

\section{RFID System Model\protect\label{sec:System_Model}}

The used system model consists of two parts. The first part is a description of the communication between the RFID reader and the RFID tag according to the UHF-RFID transmission protocol, including modulation, encoding, and signal timings. This is provided in Section \ref{subsec:System_Model_EPC_Global}. The second part is a channel model, where our assumptions regarding the movement of the tags, the communication channel, as well as received noise and tag signal strength are described. This is provided in Section \ref{subsec:Chanel_Model}.

\subsection{Introduction to the EPCglobal UHF-RFID Protocol \protect\label{subsec:System_Model_EPC_Global}}

Our work assumes UHF-RFID tags that communicate with a UHF-RFID reader according to the EPCglobal Class-1 Generation-2 Protocol~\cite{Spec}, which is the most common protocol for UHF-RFID.

\subsubsection{Frequency Bands}

UHF-RFID operates in the frequency bands around $\unit[868]{MHz}$ or $\unit[915]{MHz}$, depending on the country-specific frequency regulations. 
In this work we assume the operation in the European frequency band around $\unit[868]{MHz}$.
In Section \ref{sec:influence-of-carrier-frequency}, we will also briefly discuss the differences in the performance of readers operating in the $\unit[915]{MHz}$ band, which is the UHF-RFID frequency band used in the Americas.

\subsubsection{Modulation}

The tags communicate to the reader by reflecting a continuous
wave that is sent by the reader during the whole duration of the communication.
The tag data is mo\-dulated onto the continuous wave by switching the reflectivity factor at the antenna, so that the wave is reflected with different phase shift or amplitude, depending on the transmitted symbol~\cite{finkenzeller2015rfid}. 
The EPCglobal protocol defines Amplitude Shift Keying (ASK) and Phase Shift Keying (PSK) for the communication from the tag to the reader. 
The decision between the two is not up to the reader, it is determined by the tag manufacturer.

UHF-RFID uses binary modulation, so the tag has two different backscatter states in which the carrier signal is reflected differently, which we will later name "zero mode" and "one mode". For PSK, these differ by a phase-shift of the reflected signal of ideally 180°.
For ASK, the two states differ by the amplitude of the reflected signal. The reflected signal for ASK in this case looks like an On-Off keying signal~\cite{Proakis2007} as the tag in this case implements two states that can be called the ``reflect state'', in which the carrier signal is reflected, and ``absorb state'', in which the carrier signal is absorbed and not reflected.

\subsubsection{Medium Access}
The communication between the reader and the tags is organized based on the frame-slotted ALOHA principle~\cite{Proakis2007}.
According to the EPCglobal standard, an ALOHA frame is named inventory round.
It is set up by the reader sending a so-called Query command that sets up the communication and specifies parameters like the Backscatter Link Frequency (BLF), data encoding, and the number of ALOHA slots in which the tags then respond.
Each tag then randomly chooses one slot to reply.
The communication for a given tag is successful, if it is the only tag responding in a specific slot.

\begin{figure}[t]
\centering{}\includegraphics[width=3.5in]{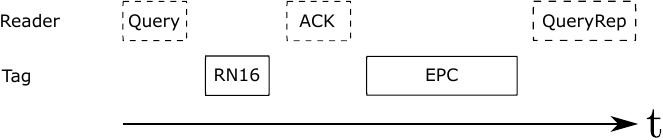}
\caption{Organization of the communication between reader (messages with dashed
border) and one tag (messages with solid border) \cite{Spec}. The
reader starts the inventory round with a Query command. In a successful
slot, the tag then replies by backscattering an RN16. The RN16 is
acknowledged by the reader\textquoteright s ACK command which commands
the tag to transmit its EPC. This can be followed by a Query Repeat
command (QueryRep) which indicates the next slot in which another
tag can reply.}
\label{fig:ErfolgreicherSlotFig}
\end{figure}

Fig.~\ref{fig:ErfolgreicherSlotFig} shows the communication between the reader and the tag during a successful
slot with a single tag response. For a successful inventory, the tag
first backscatters an RN16 (16 bit random number). 
After successfully receiving this number, the reader acknowledges the RN16 with a so-called
ACK (acknowledge) command, which then commands the tag to backscatter
its 96-bit Electronic Product Code (EPC). 
For Doppler-based motion
detection, these two tag signals (RN16 and EPC) can be used to
estimate the Doppler shift, and for this reason we also derive the bounds for the Doppler shift estimation based on these two distinct tag signal parts.
As we will see later, this provides a significant gain compared to classical methods.

\subsubsection{Backscatter-Link Frequency (BLF) and Tag Uplink Data Encoding}
The Backscatter-Link Frequency (BLF) defines the rate at which a tag modulates its uplink data.
It ranges from $\unit[40]{kHz}$ to $\unit[640]{kHz}$, and is configured by the reader in the Query command at the start of an inventory round.
Further, the tags support the FM0, Miller-2, Miller-4, and Miller-8 encoding schemes.
The used encoding scheme is also signaled in the Query command.
FM0 is principally an orthogonal modulation~\cite{Proakis2007} based on the aforementioned backscatter modulation.
The Miller modes add an additional signal spreading by the spreading factor $M$ given by:
\[
M=\left\{ \begin{array}{c}
1\quad\mathtt{\mathit{\mathrm{for}}}\quad\mathrm{\textrm{FM0}}\:\;\;\;\;\\
2\quad\mathrm{for}\quad\textrm{Miller-2}\\
4\quad\mathit{\mathrm{for}}\quad\mathrm{\textrm{Miller-4}}\\
8\quad\mathit{\mathrm{for}}\quad\mathrm{\textrm{Miller-8}}
\end{array}\right..
\]
This spreading increases the overall signal duration by the factor $M$.

\subsubsection{Tag Uplink Signal Duration}
\label{sec:tag_signal_durations}

The duration of tag signals in the UHF-RFID standard is crucial for calculating bounds. 
Table~\ref{tab:signal_times_EPCglobal} lists signal durations for the minimum and maximum BLFs ($\unit[40]{kHz}$ and $\unit[640]{kHz}$) across different encoding schemes specified in \cite{Spec}. 
Intermediate values are possible if the reader adjusts the BLF accordingly.
The durations assume the reader has instructed tags to use the long preamble with a pilot tone (TRext = 1 in the Query command). While various EPC lengths exist, this paper considers the 96-bit EPC as the default, as it is the most common.

\begin{table}[tbh]
\caption{Selected Signal Durations for the EPCglobal Standard}
\subfloat[FM0 encoding]{%
\begin{tabular}{|c|c|c|}
\hline 
$BLF$ & $\unit[40]{kHz}$ & $\unit[640]{kHz}$\tabularnewline
\hline 
\hline 
$T_{RN16}$ & $\unit[0.875]{ms}$ & $\unit[0.0547]{ms}$\tabularnewline
\hline 
$T_{EPC}$ & $\unit[3.275]{ms}$ & $\unit[0.2047]{ms}$\tabularnewline
\hline 
\end{tabular}

}\subfloat[Miller-2 encoding]{%
\begin{tabular}{|c|c|c|}
\hline 
$BLF$ & $\unit[40]{kHz}$ & $\unit[640]{kHz}$\tabularnewline
\hline 
\hline 
$T_{RN16}$ & $\unit[1.95]{ms}$ & $\unit[0.122]{ms}$\tabularnewline
\hline 
$T_{EPC}$ & $\unit[6.75]{ms}$ & $\unit[0.422]{ms}$\tabularnewline
\hline 
\end{tabular}

}

\subfloat[Miller-4 encoding]{%
\begin{tabular}{|c|c|c|}
\hline 
$BLF$ & $\unit[40]{kHz}$ & $\unit[640]{kHz}$\tabularnewline
\hline 
\hline 
$T_{RN16}$ & $\unit[3.90]{ms}$ & $\unit[0.244]{ms}$\tabularnewline
\hline 
$T_{EPC}$ & $\unit[13.50]{ms}$ & $\unit[0.843]{ms}$\tabularnewline
\hline 
\end{tabular}

}\subfloat[Miller-8 encoding]{%
\begin{tabular}{|c|c|c|}
\hline 
$BLF$ & $\unit[40]{kHz}$ & $\unit[640]{kHz}$\tabularnewline
\hline 
\hline 
$T_{RN16}$ & $\unit[7.80]{ms}$ & $\unit[0.488]{ms}$\tabularnewline
\hline 
$T_{EPC}$ & $\unit[27.00]{ms}$ & $\unit[1.688]{ms}$\tabularnewline
\hline 
\end{tabular}

}

\label{tab:signal_times_EPCglobal}
\end{table}
The duration of the pause between RN16 and EPC is not strictly defined, it depends on various reader parameters, some of which can be adjusted within a specified range.  
For simplicity, we use example pause durations and assume that both the pause length and signal duration scale linearly with the BLF within the following range:
\[
T_{pause}=\left\{ \begin{array}{ll}
\unit[1.4]{ms} & \mathit{\mathsf{\mathrm{for}}}\quad\unit[BLF=40]{kHz}\\
\unit[0.2]{ms} & \mathit{\mathrm{for}}\unit[\quad BLF=640]{kHz}
\end{array}\right.
\]

\subsection{Channel Model \protect\label{subsec:Chanel_Model}}

\begin{figure}[t]
\centering{}\includegraphics[width=3.5in]{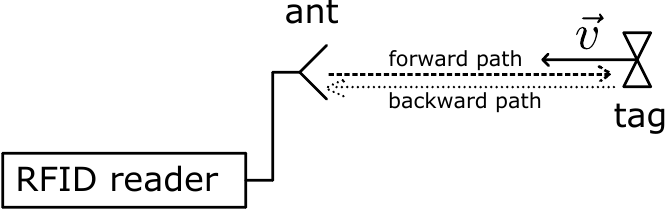}
\caption{The channel is modeled by a static RFID reader that communicates with
a tag while the tag is moving with constant speed $\vec{v}$ into
the direction of the reader antenna. The forward path is depicted
by the dashed arrow while the backward path is depicted by the dotted
arrow.}
\label{fig:channel_model}
\end{figure}
The typical reading range of UHF-RFID is only a few meters.  
As a result, the channel's delay spread will be relatively short.  
Combined with the narrow bandwidth of UHF-RFID systems, this permits a simplified, non-frequency-selective fading model.  
Accordingly, the effect of the channel $h(t)$ can be modeled as a multiplication rather than a convolution operation~\cite{molisch2012wireless}.  
We further assume a strong line-of-sight (LoS) component.  
Given the short communication duration, the channel attenuation due to the varying distance between tag and reader can be considered constant during a single communication exchange.  
Therefore, we assume the channel to be time-invariant, except for the Doppler effect.

\subsubsection{Received Tag Signal}
In the following, we examine the
signals in complex baseband representation~\cite{Proakis2007}. 
We assume perfect carrier suppression in the RFID reader's receiver circuit and no phase noise in the received tag signal.
Then, the tag signal received by the reader is given by

\begin{equation}
s(t)=a(t)h_{Doppler}(t)+n(t)\label{eq:channel_model},
\end{equation}
where $a(t)$ is the amplitude of the received tag signal, $h_{Doppler}(t)$ is a channel coefficient that describes the influence of the Doppler shift on the received tag signal, and $n(t)$ is complex additive white Gaussian noise.

For PSK, the tag is in reflect mode all the time during data transmission and only changes the phase of the backscattered signal. 
Therefore, the amplitude for PSK is given by
\begin{equation}
a_{PSK}(t)=\left\{ \begin{array}{ll}
\sqrt{P_S} & \mathit{\mathsf{\mathrm{in\:``zero\:state``}}}\\
\sqrt{P_S}\exp(j \pi)=-\sqrt{P_S} & \mathit{\mathsf{\mathrm{in\:``one\:state``}}}
\end{array}\right.,
\end{equation}
where $P_S$ is the average signal power of the received tag signal.

For ASK, the amplitude in the "one state" is $\sqrt{P_{refl}}$, where $P_{refl}$  is the power of the reflected tag signal that the reader receives during this time, and zero during the "zero state". 
Due to the FM0 or Miller encoding, the tag is exactly half of the time in the "zero state" and half of the time in "one state" during data transmission. Therefore, the average signal power received during the transmission of the tag signal is $P_S=P_{refl}/2$. The amplitude for tag signals with ASK modulation is given by
\begin{equation}
a_{ASK}(t)=\left\{ \begin{array}{ll}
0 & \mathit{\mathsf{\mathrm{in\:``zero\:state``}}}\\
\sqrt{P_{refl}}=\sqrt{2 P_S} & \mathit{\mathsf{\mathrm{in\:``one\:state``}}}
\end{array}\right..
\end{equation}

\subsubsection{Doppler Shift}

The influence of the Doppler effect on the channel is a frequency shift that can be described by
\begin{equation}
h_{Doppler}(t)=\exp\left(-j 2\pi f_{D}t\right)\label{eq:h_doppler},
\end{equation}
where $f_{D}$ is the experienced Doppler shift.
In general, a Doppler shift is calculated by $f_D=v_{rel}/\lambda_{c}$, where $v_{rel}$ denotes
the velocity of the moving object relative to the receiver antenna, and $\lambda_{c}$ the wavelength
of the continuous wave (CW) (cf. \cite[(5.57)]{Rappaport}). 

Our RFID scenario corresponds to a mono-static radar-like case, where the Doppler shift is experienced twice: Once in the forward channel, and once in the backward channel.
This has to be considered by multiplying the Doppler shift by two. Then, using $\lambda_{c}=c/f_{c}$, where $c$ is the speed
of light, and $f_{c}$ the carrier frequency of the transmitted signal (in our case the carrier signal radiated by the RFID reader), yields:
\begin{equation}
f_{D}=2 \dfrac{v_{rel}}{c}f_{c}.\label{eq:doppler_shift_formula}
\end{equation}

\subsubsection{Movement Direction}

To simplify calculations, we assume the tag moves directly toward or away from the reader antenna, as illustrated in Fig.~\ref{fig:channel_model}. 
In this case, the tag's relative speed to the antenna equals its actual speed, i.e., \( v_{rel} = v \). 
This avoids the need for an additional term to account for movement direction angles.
Further, we assume that the tag speed $v$ is known and constant, which is typically the case for a conveyor belt scenario.

Using these assumptions together with (\ref{eq:doppler_shift_formula}) in (\ref{eq:h_doppler}) yields
\begin{equation}
h_{Doppler}(t)=\exp\left(-j 4\pi \dfrac{ v }{c} f_{c} t\right)\label{eq:h_doppler_final}.
\end{equation}

\subsubsection{Noise Level and Noise Figure\label{sec:NoiseFigure}}
Generally, all terrestrial receivers are affected by thermal noise with a noise spectral density of $\unit[-174]{dBm\text{-}Hz}$~\cite{molisch2012wireless}.
Furthermore, we must consider receiver imperfections, which result in additional noise as, e.g., given by the noise figure $NF$ of the receiver~\cite{molisch2012wireless}.
Hence, the  one-sided noise power spectral density is given by 
\begin{equation}
N_{0}=\unit[-174]{dBm\text{-}Hz}+NF,\label{eq:N_0}
\end{equation}
where $NF$ in dB denotes the noise figure of the reader.

The knowledge of $N_0$ is important for the interpretation of the achievable estimation precision.
However, the noise figure is not mentioned in the data sheets of UHF-RFID readers.
Therefore, we try to estimate the noise figure of a state-of-the-art RFID reader (Impinj in R700) in Annex~\ref{Sec:NoiseFigureEst}.
The estimated noise spectral density is $N_{0}=\unit[-148.6]{dBm\text{-}Hz}$, corresponding to a noise figure of $NF=\unit[25.4]{dB}$.
This seems quite high, but we have to take into consideration that the reader has to work in full-duplex mode, i.e., it is transmitting and receiving on the very same frequency.

\subsubsection{Receiver Sensitivity and Received Tag Power\label{sec:Realistic-PSN0}}
To interpret the theoretical bounds meaningfully, it is important to compare them against the reader's sensitivity for data decoding -- i.e., the minimum required received power $P_S$ for successful decoding.  
The most sensitive mode of the reader under consideration is Mode 290 ($BLF = \unit[160]{kHz}$, Miller-8), which requires $P_S = \unit[-95.8]{dBm}$ to achieve a bit-error-rate (BER) below $10^{-3}$.  
Accordingly, this mode will be used as the sensitivity threshold in the subsequent evaluations.

In this work, we use the ratio between the signal power $P_S$, and the one-sided noise spectral density $N_0$ as a measure of the tag signal quality.
Unlike the classical signal-to-noise ratio (SNR), this metric is independent of the noise bandwidth and thus not affected by the receiver bandwidth~\cite{Proakis2007}.
As a result, it serves as a hardware-independent indicator of signal quality.  
For the Impinj R700 in Mode 290, the reference value for the weakest reliably received tag signal is $\unit[P_S / N_0 = 52.8]{dB\text{-}Hz}$.

\section{Required Estimation Precision for Reliable Motion Detection\protect\label{sec:Minimum_Required_Estimation_Precision}}

This section examines the required Doppler shift estimation precision for reliably distinguishing tags moving with a given velocity $v$ from stationary ones. 
Assuming the estimation error follows a normal distribution — justified by the maximum likelihood estimation (MLE) approach, which converges to a normal distribution~\cite{Proakis2007} — the precision is fully characterized by the estimation variance $\sigma^2$.

\begin{figure}[t]
\centering{}\includegraphics[width=3.5in]{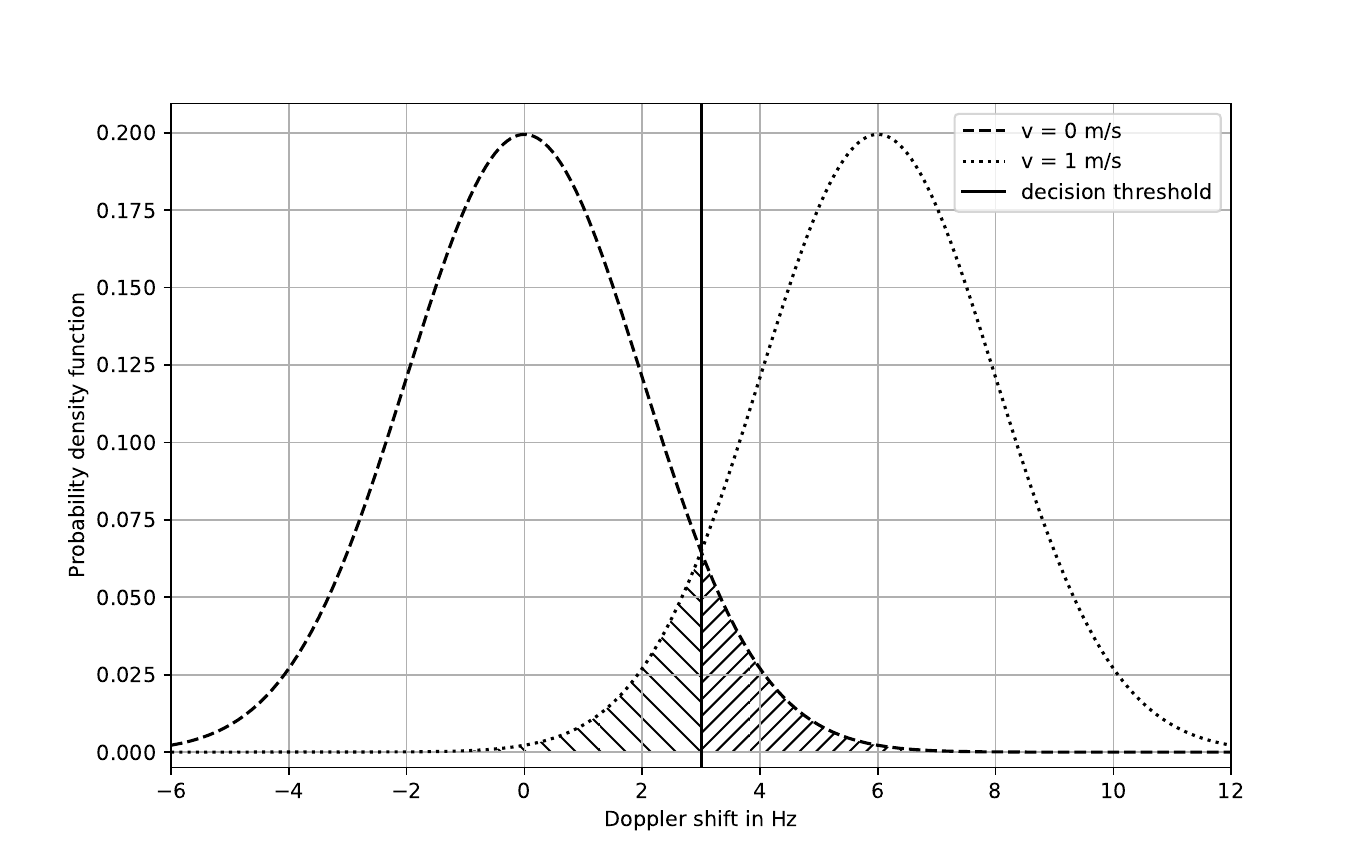}
\caption{Probability density functions of the estimated Doppler shifts for
two tags: One stationary tag (Doppler shift $\unit[0]{Hz}$) and one
that is moving with a velocity of $\unit[1]{m/s}$ (leads to a Doppler
shift of $\unit[6]{Hz}$ at a carrier frequency of $\unit[900]{MHz}$).
The solid line shows the decision threshold at $\unit[3]{Hz},$ the
two areas are the areas that cause estimation errors (\textbackslash\textbackslash\textbackslash -hatched:
moving tag detected as stationary; ///-hatched: static tag detected
as moving).}
\label{fig:Gauss_curves}
\end{figure}

We consider two types of tags to be distinguished: a stationary tag with a velocity of $\unit[0]{m/s}$ and a moving tag with $\unit[1]{m/s}$. Fig.~\ref{fig:Gauss_curves} shows the probability density functions for the estimated velocity given a variance $\sigma^2$.
For simplicity, we assume a carrier frequency of $\unit[900]{MHz}$.
In this case, the Doppler effect shifts the received signal by exactly $\unit[6]{Hz}$ for moving tags, while stationary tags experience no shift.
To classify the tags, we use a decision threshold at $\unit[3]{Hz}$: Tags with an estimated Doppler shift below $\unit[3]{Hz}$ are classified as stationary, while those above $\unit[3]{Hz}$ are classified as moving.

Fig.~\ref{fig:Gauss_curves} shows the probability density functions of the estimated frequency shifts for the moving and the stationary tag, and the decision threshold at $\unit[3]{Hz}$. Assignment errors occur in the hatched overlapping regions of both curves. 
Values from the dashed $\unit[0]{m/s}$ curve exceeding the threshold (///-hatched area) are misclassified as moving tags, despite being stationary.
Conversely, values from the dotted $\unit[1]{m/s}$ curve falling below the threshold (\textbackslash\textbackslash\textbackslash-hatched area) are incorrectly detected as stationary.
The extent of the overlap -- and, thus, the error-prone region -- is determined by the variance $\sigma^2$ and the velocity of the tag, $v$.
Hence, it is much simpler to classify stationary and moving tags if the velocity is higher.
However, to minimize classification errors for a given velocity, $\sigma^2$ has to be reduced.
Based on this, we can compute the maximum allowable variance for a given classification error rate $P_{err}$, assuming equal probabilities for stationary and moving tags.
\begin{thm}
\label{thm:sigma_max}
To distinguish between a tag moving at speed $v$ and a stationary tag using Doppler shift estimation, with an error probability $P_{\text{err}}$, the maximum tolerable variance of the Doppler shift estimate is given by:
\begin{align}
\sigma_{max}^{2} & =\dfrac{v^{2}f_{c}^{2}}{2c^{2}\left(\textrm{\ensuremath{\erf^{-1}}}\left(1-2P_{err}\right)\right)^{2}}\label{eq:required_estimation_variance-general}
\end{align}
\end{thm}
\begin{IEEEproof}
The proof is given in Appendix \ref{sec:Appendix-required_estimation_precision}.
\end{IEEEproof}

\begin{figure}[t!]
\centering{}\includegraphics[width=3.5in]{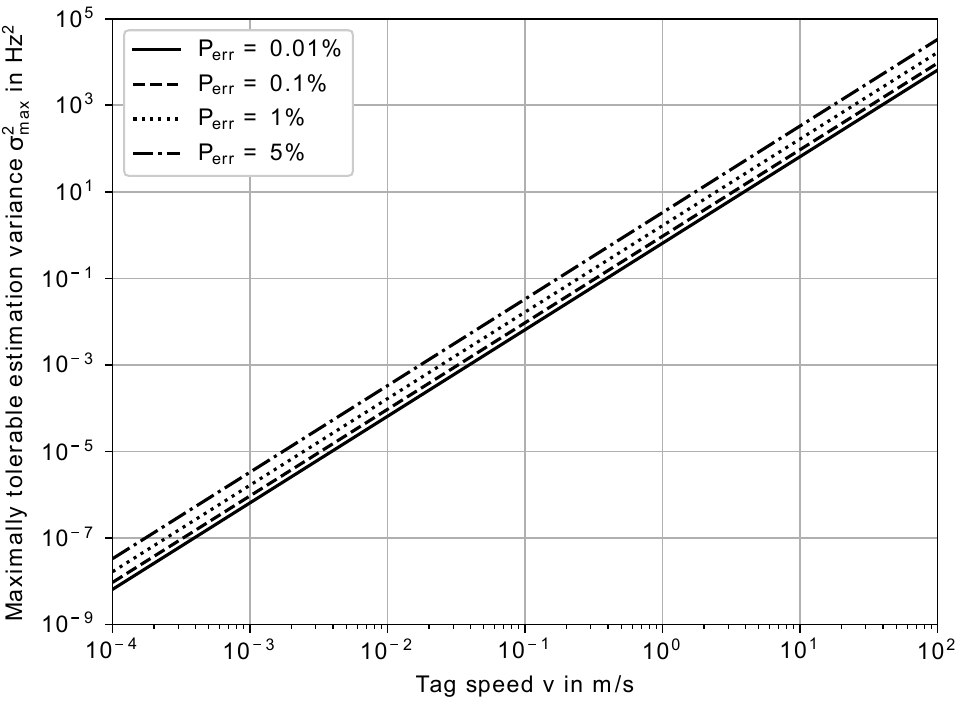}
\caption{This figure shows the maximally tolerable estimation variance, $\sigma_{max}^{2}$,
as a function of the tag speed, $v$, for various pursued error probabilities,
$P_{err}$, and carrier frequency $\unit[f_{c}=868]{MHz}$. }
\label{fig:tolerable-estimation-variance_over_tag-speed}
\end{figure}

Fig.~\ref{fig:tolerable-estimation-variance_over_tag-speed} shows the maximum estimation variance $\sigma_{\max}^{2}$ as a function of the speed of the tag for different probabilities of classification error. The variance increases quadratically with the speed of the tag.
The figure also illustrates the impact of the error rate $P_{err}$. 
A higher $P_{err}$ allows for greater overlap between the two Gaussian curves, meaning that they can be wider, and thus the tolerable estimation variance increases. 
However, the effect of $P_{err}$ on the precision of the estimation is relatively small: Reducing the tolerable error rate from 5\% to 0.1\% only doubles the smallest distinguishable tag speed.
\section{Achievable Estimation Precision for UHF-RFID\protect\label{sec:MCRBSection}}

In the previous section, we derived the maximum variance $\sigma_{\max}^{2}$ for a given classification error rate $P_{err}$.
Now, we investigate which estimation variance can be achieved by a Doppler shift estimator for given tag signals. To achieve this, we derive the theoretical bound for the estimation variance as a function of $P_{S}/N_{0}$.

\subsection{Derivation of the Bound}

The achievable estimation variance can be predicted theoretically by the Cramér-Rao lower bound (CRLB), which provides a lower limit on the variance of unbiased estimators~\cite{candy2016bayesian}.
However, due to its complexity, we use the Modified Cramér-Rao Bound (MCRB)~\cite{MCRBAndrea} instead. 
The MCRB is easier to calculate, but not necessarily tight, meaning that the results may be overly optimistic. 
However, we later present simulation results that fully overlap with the MCRB, confirming that the derived bounds are indeed tight.

\begin{figure}[t]
\centering{}\includegraphics[width=2.5in]{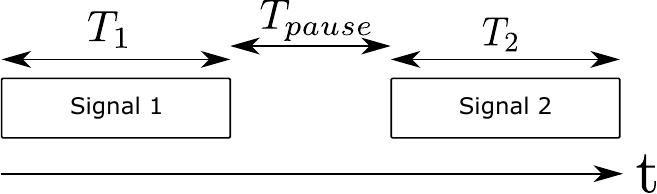}
\caption{Tag reply timing: $T_{1}$ denotes the length of the first packet,
$T_{2}$ the length of the second packet, $T_{pause}$ denotes the
length of the pause between the two signals.}
\label{fig:signal-timing_two_signal-parts}
\end{figure}

\begin{figure}[t!]
\centering{}\includegraphics[width=3.5in]{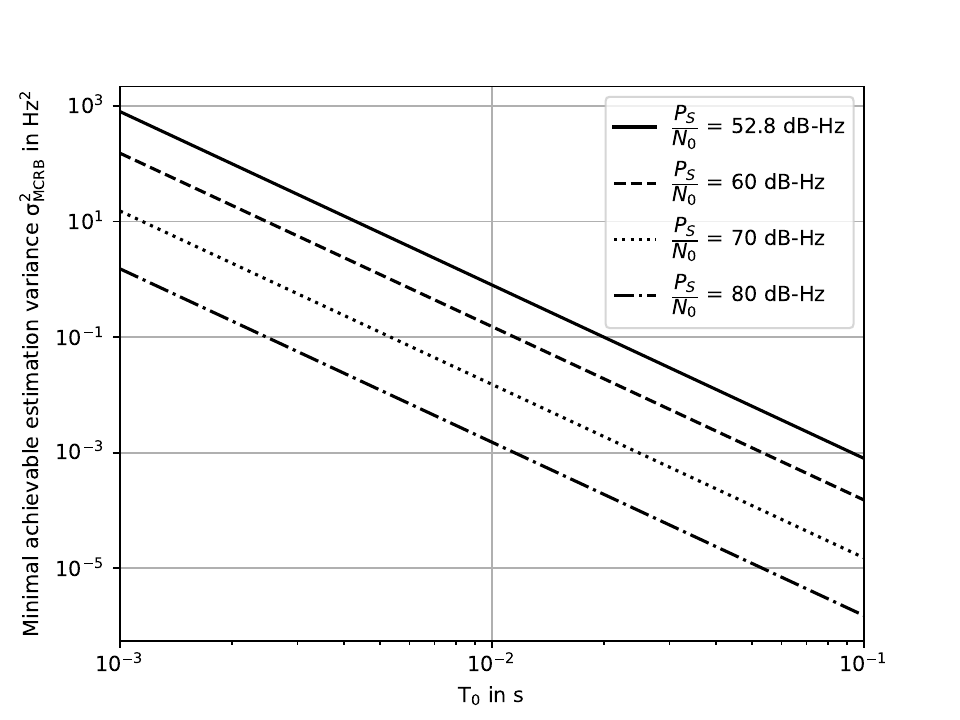}
\caption{Bound for the estimation variance for estimating the frequency of
one single signal part over the signal duration $T_{0}$.}
\label{fig:MCRB_vs_T_one_signal}
\end{figure}

As shown in Fig.~\ref{fig:MCRB_vs_T_one_signal}, each successful tag response consists of two parts -- i.e. the RN16 and the EPC -- separated by a pause. To consider that, we calculate the bound for one signal and also for tag replies that consist of two signal parts. In the case of one signal, we denote the length of the signal as $T_0$. In the case of two signal parts, the durations are given as shown in Fig.~\ref{fig:signal-timing_two_signal-parts}. $T_{1}$ is the duration of the first signal (e.g. transmission of the RN16), $T_{2}$ is the duration of the second signal (e.g. transmission of the EPC) and $T_{pause}$ is the duration of the pause between the two signal parts.

\begin{thm}
\label{thm:MCRB1} The minimum achievable estimation variance for the Doppler shift of a UHF-RFID signal is given by:

\begin{equation}
\sigma_{MCRB}^{2}=\dfrac{3}{ 2 \pi^{2}C_T}\frac{N_{0}}{P_{S}},\label{eq:MCRB1_with_modulation_loss_simplified_new}
\end{equation}
where $C_T$ is:
\begin{itemize}
    \item for Doppler shift estimation based on one tag signal:
\begin{equation}
C_{T}=  T_0^3 \label{eq:CT1}
\end{equation}
    \item for Doppler shift estimation based on two tag signals:
\begin{equation}
C_{T}= (T_1 + T_2)^3 +  12 T_1 T_2 T_{pause}\dfrac{ (T_1 + T_2 + T_{pause})}{T_1 + T_2}. \label{eq:CT2}
\end{equation}
\end{itemize}

\end{thm}
\begin{IEEEproof}
The proof of Theorem \ref{thm:MCRB1} is given in Appendix \ref{sec:Appendix-single_tag_response}.
\end{IEEEproof} 

It is worth noting that the Modified Cramér-Rao Bound assumes that the received data symbols are known~\cite{MCRBAndrea}.
The bound is therefore only valid for tag signals that can be decoded correctly by the reader. 
However, in applications where the Doppler shifts of RFID tag signals are estimated, we can assume that only the Doppler shifts of correctly decoded tag signals are of interest.
For non-decodable tag signals, the Doppler shift estimate could not be assigned to any object and would therefore be useless.

\subsection{Discussion of the Bound}

Fig.~\ref{fig:MCRB_vs_T_one_signal} shows the $\sigma_{MCRB}^{2}$ as a function of $T_{0}$ for various $P_{S}/N_{0}$. 
The estimation variance decreases with $T_0$ following a cubic exponent, while the signal strength $P_{S}$ influences the variance linearly. 

\begin{figure}[t!]
\centering{}\includegraphics[width=3.5in]{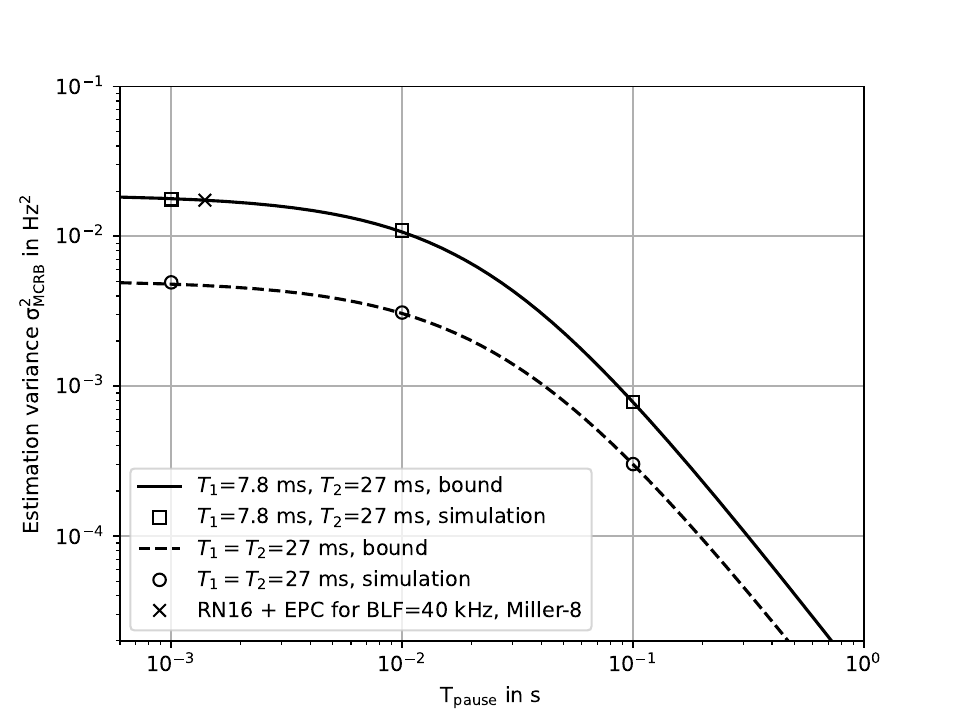}
\caption{Bound for the estimation variance for estimating the frequency of
two signal parts of length $T_1$ and $T_2$
and a pause in between
of length $T_{pause}$ for $\unit[P_{S}/N_{0}=52.8]{dB\text{-}Hz}$.  $\unit[7.8]{ms}$ is the duration of an RN16 signal with $BLF=\unit[40]{kHz}$ and Miller-8 encoding, while $\unit[27]{ms}$ is the duration of an EPC signal in the same case. The "x"-mark shows the performance for an RN16 and EPC signal with $T_{pause}=\unit[1.4]{ms}$, which is the correct bound for a tag signal for $BLF=\unit[40]{kHz}$ and Miller-8 encoding.}
\label{fig:MCRB_vs_T_pause_two_signals}
\end{figure}

Fig.~\ref{fig:MCRB_vs_T_pause_two_signals} illustrates the impact of signal lengths from two tag signal parts on the estimation variance.
The variance \( \sigma_{MCRB}^{2} \) is plotted against the pause duration \( T_{pause} \) between the RN16 and the EPC. 
The results show that estimation variance decreases as $T_{pause}$ increases, and also decreases with longer signal parts.
The solid curve represents the bound for two signal parts corresponding to an RN16 and an EPC signal at $BLF = \unit[40]{kHz}$ with Miller-8 encoding, resulting in the maximum possible signal length.
The correct pause length for this configuration, $T_{pause} = \unit[1.4]{ms}$, is marked with an "x".

The dashed curve represents the bound for the theoretical case where both signal parts match the length of an EPC signal at $BLF = \unit[40]{kHz}$ with Miller-8 encoding. This curve demonstrates that high estimation precision can be achieved when reading the tag twice with a pause between the reads.
However, ambiguities can arise in frequency estimation if two short signals are separated by a pause significantly longer than the signal durations. This phenomenon is not accounted for by the Cramér-Rao bound and is beyond the scope of this paper.

\subsection{Simulation Results}
Fig.~\ref{fig:MCRB_vs_T_one_signal} also includes results from Monte Carlo simulations. 
In these simulations, random signals with the given timings were generated, incorporating a frequency shift, random tag signal modulation, and additive white Gaussian noise, ensuring $P_{S}/N_{0} = \unit[52.8]{dB\text{-}Hz}$ for a realstic reader. 
The frequency estimation using the standard FFT algorithm~\cite{rife1974single} achieved a variance that precisely matched the derived MCRB. 
This confirms the existence of an optimal frequency estimator for this case and demonstrates that the MCRB is tight.
In order to reach the bound, a modulation-specific pre-processing of the data is required for the two possible backscatter modes of the tag.

For ASK (i.e., On-Off Keying), the received signal was set to zero whenever the tag was in the "zero state." This adjustment is only possible after successfully decoding the signal, resulting in reduced received noise.
Without this step, the Doppler shift estimator's performance would degrade by $\unit[3]{dB}$ compared to the bound.

For PSK, the tag signal's phase was shifted by $\pi$ during the "one mode" before Doppler shift estimation, effectively canceling the modulation. This step is only possible after successfully decoding the signal.

\section{Bounds for the Motion Detection\protect\label{sec:Boundaries-of-the-Motion-Detection}}

This section examines the tag speeds and signal strengths required for reliable motion detection. 
The theoretical bound is derived from the required estimation reliability in Section~\ref{sec:Minimum_Required_Estimation_Precision} and the estimation precision bound in Section~\ref{sec:MCRBSection}.  

The motion detection bound is presented in Section~\ref{sec:motion-detection-bound-subsection}, while Section~\ref{sec:limits-in-optimal-conditions} explores theoretical limits under optimal conditions and the impact of key parameters. 
Finally, Section~\ref{sec:limits-in-realistic-conditions} assesses practical limitations, considering the influence of RFID reader receiver chains.

\subsection{Theoretical Bound\label{sec:motion-detection-bound-subsection}}

\begin{thm}
The minimum tag speed at which Doppler-based
motion detection is possible with error probability $P_{err}$
is given by
\begin{equation}
v_{min}=   \dfrac{c \; \textrm{\ensuremath{\erf^{-1}}}(1-2P_{err})}{\pi f_c}    \sqrt{ \dfrac{3 }{C_T}\dfrac{N_0}{P_S}}, \label{eq:motion_detection_bound}
\end{equation}
where $C_T$ is given by (\ref{eq:CT1}) for Doppler shift estimation based on a single tag signal, and by (\ref{eq:CT2}) for Doppler shift estimation based on two tag signals.
\end{thm}
\begin{IEEEproof}
Reliable motion detection is possible when the estimation
variance achieved, given by (\ref{eq:MCRB1_with_modulation_loss_simplified_new}), is not
greater than the maximum tolerable estimation variance, given by
(\ref{eq:required_estimation_variance-general}). Setting $\sigma_{max}^{2}=\sigma_{MCRB}^{2}$
with (\ref{eq:MCRB1_with_modulation_loss_simplified_new}) and (\ref{eq:required_estimation_variance-general})
and solving the result for $v$ yields (\ref{eq:motion_detection_bound}).
\end{IEEEproof}

\subsection{Limits Under Ideal Conditions\label{sec:limits-in-optimal-conditions}}
\subsubsection{Relation between $P_{S}/N_{0}$ and $v_{min}$}
\begin{figure}[t!]
\centering{}\includegraphics[width=3.5in]{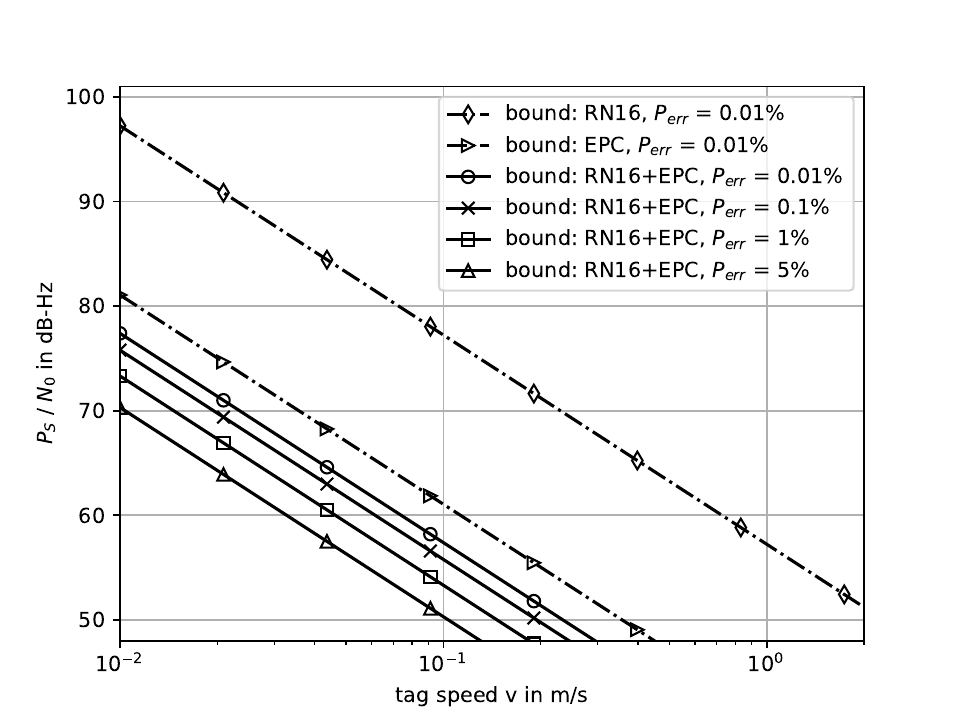}
\caption{Required $P_{S}/N_{0}$ for a reliable motion detection over the tag
speed for different detection error probabilities, $P_{err}$, for
Miller-8 encoding and $BLF=\unit[40]{kHz}$.}
\label{fig:PSN0_over_tag-speed_various-Perr}
\end{figure}

Fig. \ref{fig:PSN0_over_tag-speed_various-Perr} shows the required $P_{S}/N_{0}$ for reliable motion detection as a function of the tag speed $v$. 
It is obtained by solving (\ref{eq:motion_detection_bound}) for $P_{S}/N_{0}$ and demonstrates a quadratic decrease in the required $P_{S}/N_{0}$ as $v$ increases.
A higher $v$ increases the separation between the Gaussian curves in Fig.~\ref{fig:Gauss_curves}, allowing for wider distributions while maintaining the same overlap. This increases the tolerable estimation variance, enabling accurate estimation even at lower $P_{S}/N_{0}$.

The curves in Fig.~\ref{fig:PSN0_over_tag-speed_various-Perr} represent Miller-8 encoding with $BLF = \unit[40]{kHz}$, producing the longest possible signals under the EPCglobal standard~\cite{Spec}.  
As a result, these curves define the absolute lower limit for commercial UHF-RFID systems.

\subsubsection{Single Signal vs. Combined Signals}
Fig.~\ref{fig:PSN0_over_tag-speed_various-Perr} also compares the use of only the RN16 signal, only the EPC signal, and both signals combined.  
Using the EPC signal instead of the RN16 signal provides a significant gain, reducing $v_{min}$ by a factor of $6.4$ or lowering $P_S/N_0$ by $\unit[16]{dB}$. Additionally, combining both signals further improves performance, decreasing $v_{min}$ by a factor of $1.5$ or reducing $P_S/N_0$ by $\unit[3.6]{dB}$.


\subsubsection{Influence of $P_{err}$}
Fig.~\ref{fig:PSN0_over_tag-speed_various-Perr} also shows that the required $P_{S}/N_{0}$ decreases as the tolerated error probability $P_{err}$ increases.
A higher $P_{err}$ allows for greater overlap between the Gaussian curves in Fig.~\ref{fig:Gauss_curves}, enabling wider distributions and thus a lower acceptable $P_{S}/N_{0}$.

\subsubsection{Influence of $BLF$ and Data Encoding Scheme}
\begin{figure}[t!]
\centering{}\includegraphics[width=3.5in]{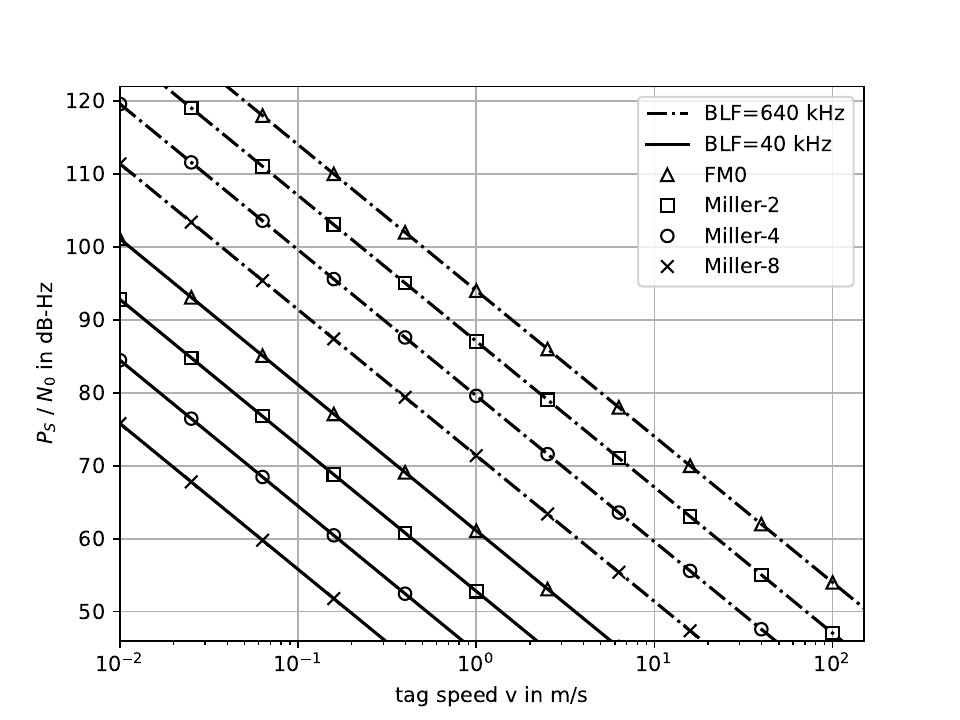}
\caption{Required $P_{S}/N_{0}$ for a reliable motion detection over the tag
speed for different encodings and $BLF$s, for $P_{err}=0.1\%$.} 
\label{fig:PSN0_over_tag-speed_various-Mod-and-BLF}
\end{figure}

Fig.~\ref{fig:PSN0_over_tag-speed_various-Mod-and-BLF} presents the bound for various encodings and for $BLF = \unit[640]{kHz}$, the highest BLF allowed by the UHF-RFID standard. 
The bounds are shown for motion detection using both RN16 and EPC, with a target error probability of $P_{err} = 0.1\%$.  

For encodings other than Miller-8, the required $P_{S}/N_{0}$ increases due to shorter tag signals. 
Miller-8 is the optimal choice for maximizing estimation precision, with lower $BLF$ values yielding better results.


\subsubsection{Influence of the Frequency Band\label{sec:influence-of-carrier-frequency}}

The usable frequency bands differ slightly around the world.
The typical frequency bands for UHF-RFID are the $\unit[868]{MHz}$ and the
 $\unit[915]{MHz}$ frequency bands.

Equation~(\ref{eq:motion_detection_bound}) shows that the minimum required tag speed decreases with increasing carrier frequency, following $v_{min} \sim 1/f_{c}$, as the required minimal estimation variance increases with $\sigma_{max}^{2} \sim f_{c}^{2}$.  
Thus, motion detection performs slightly better in the $\unit[915]{MHz}$ band, with $v_{min}(f_c=\unit[915]{MHz}) \approx 0.95 \times v_{min}(f_c=\unit[868]{MHz})$, i.e., an improvement of approx. 5\% for the higher frequency.

\subsection{Limits Under Realistic Conditions\label{sec:limits-in-realistic-conditions}}

\subsubsection{Influence of the Receiver Chain}

\begin{figure}[t!]
\centering{}\includegraphics[width=3.5in]{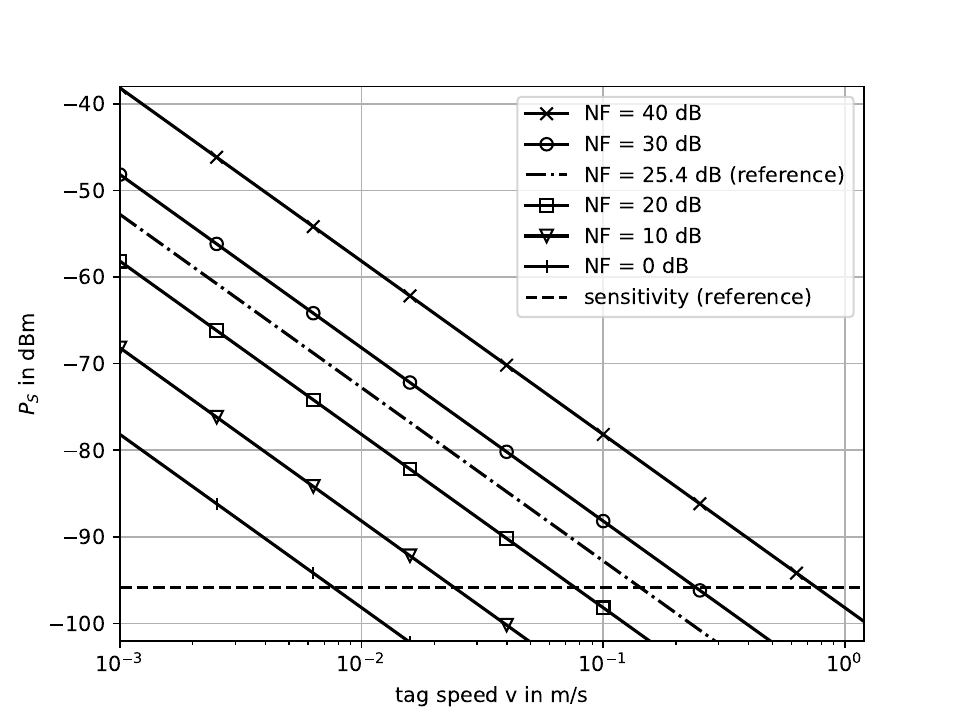}
\caption{Required tag signal power, $P_{S}$, for a reliable motion detection
over the tag speed for different noise figures, $NF$, an error rate
of $P_{err}=0.1\%$ and Miller-8 encoding with $BLF=\unit[40]{kHz}$.}
\label{fig:PS_over_tag-speed_various-N0}
\end{figure}

The receiver chain of the RFID reader is now considered by analyzing how the theoretical bounds depend on the noise figure $NF$. 
Using (\ref{eq:N_0}) in (\ref{eq:motion_detection_bound}), we determine $P_{S,min}$ as a function of the $NF$.  
The results, shown in Fig.~\ref{fig:PS_over_tag-speed_various-N0}, correspond to Miller-8 encoded tag signals with $BLF = \unit[40]{kHz}$ and $P_{err} = 10^{-3}$.

The noise figure is directly proportional to the required signal power.
Using (\ref{eq:N_0}) in (\ref{eq:motion_detection_bound}), it follows that the required tag speed scales as $\sim 1/\sqrt{NF}$. 
Thus, increasing the noise figure by $\unit[3]{dB}$ reduces the distinguishable tag speed by a factor of $\sqrt{2} \approx 1.41$.

A noise figure of $NF=\unit[0]{dB}$ corresponds to a perfect receiver without any additional degradation of the signal-to-noise ratio.
Therefore, the curves for $NF=\unit[0]{dB}$ can be
seen as the theoretical lower bound that cannot be outperformed by
any receiver. 
For reference, the dashed-dotted curve in Fig.~\ref{fig:PS_over_tag-speed_various-N0} represents the bound for a noise figure of $NF = \unit[25.4]{dB}$, corresponding to the estimated noise figure of the Impinj R700 RFID reader (cf. Sec.~\ref{sec:NoiseFigure}).  
Additionally, the horizontal dashed line indicates "Mode 290" of this RFID reader, its most sensitive configuration, which requires a minimum tag reception level of $P_S = \unit[-95.8]{dBm}$.

\subsubsection{Bounds for Real-world Reader}

\begin{figure}[t!]
\centering{}\includegraphics[width=3.5in]{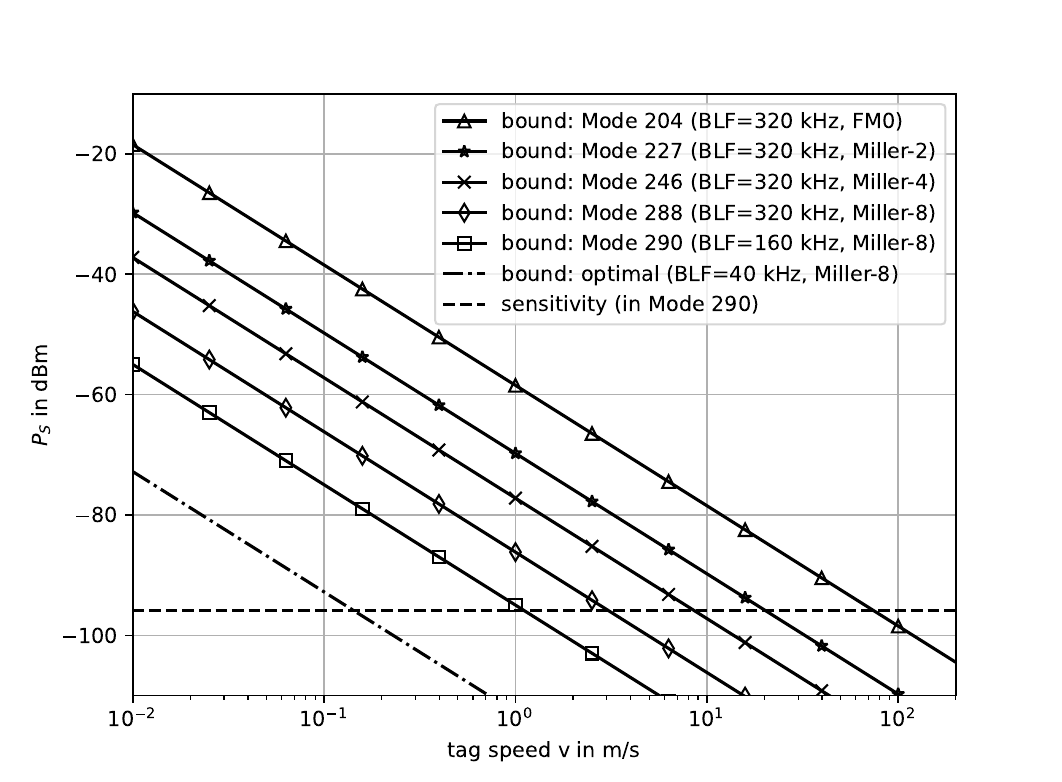}
\caption{This figure shows the bound for motion detection with $P_{err}=10^{-3}$ for some of the reader modes of the Impinj R700 RFID reader for the $\unit[868]{MHz}$ band. In addition, it also shows the bound for the optimal choice of $BLF$ and encoding as well as the minimum reception level of the Impinj R700 in Mode 290 ($BLF=\unit[160]{kHz}$, Miller-8).}
\label{fig:PS_over_tag-speed_various-Mod-and-BLF}
\end{figure}

Fig.~\ref{fig:PS_over_tag-speed_various-Mod-and-BLF} presents the bounds for various reader modes of the Impinj R700 in the $\unit[868]{MHz}$ band, assuming $N_{0} = \unit[-148.6]{dBm\text{-}Hz}$.  
The bounds are shown for the combined use of RN16 and EPC signals and for achieving $P_{err} = 10^{-3}$.

\begin{figure}[t!]
\centering{}\includegraphics[width=3.5in]{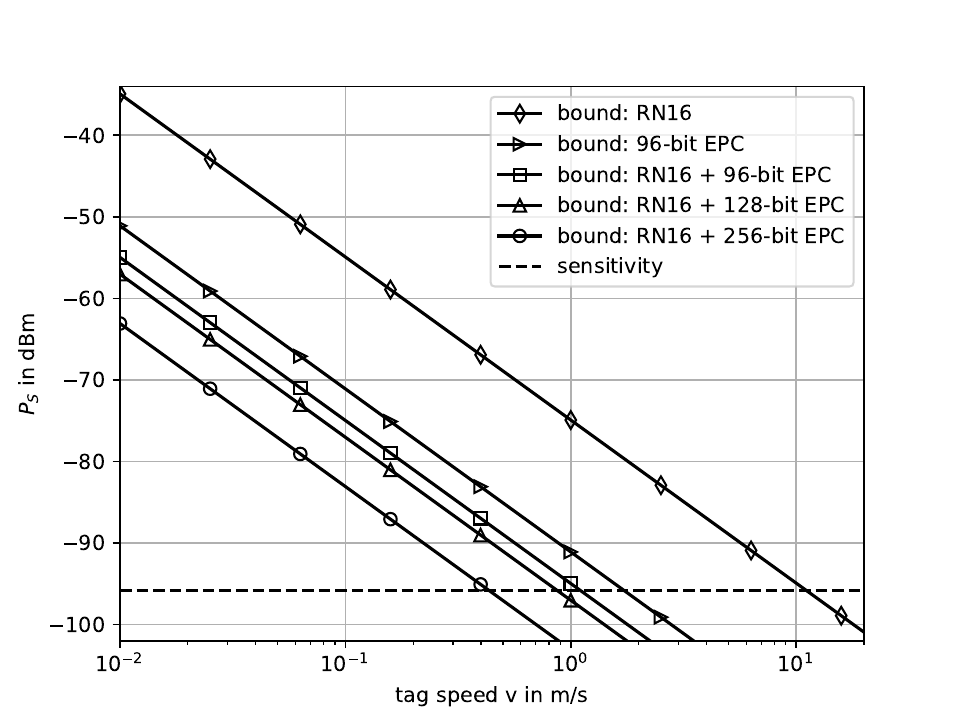}
\caption{This figure shows the bound for motion detection with $P_{err}=10^{-3}$ for the Impinj R700 in reader in Mode 290 ($BLF=\unit[160]{kHz}$, Miller-8) with $f_c=\unit[868]{MHz}$).
The horizontal dahsed line shows the minimum reception level of the Impinj R700 in Mode 204.} 
\label{fig:PS_over_tag-speed_Impinj_lengths}
\end{figure}

Mode 290 ($BLF=\unit[160]{kHz}$, Miller-8) offers the best motion detection performance, as it has the lowest data rate, and hence, the longest transmission duration. 
It requires a minimum tag reception level of $P_S = \unit[-95.8]{dBm}$ (indicated by dashed line).
This level results in a minimum tag velocity of $v_{min} = \unit[1.1]{m/s}$ for reliable motion detection.  
For stronger tag signals, $v_{min}$ decreases, e.g., to only $\unit[0.02]{m/s}$ for $P_S = \unit[-60]{dBm}$.

Mode 204 ($BLF = \unit[320]{kHz}$, FM0), which has the highest data rate, exhibits the lowest motion detection performance.  
For reference, the dashed-dotted curve represents the bound for $BLF = \unit[40]{kHz}$ with Miller-8 encoding, where performance is even better ($v_{min} = \unit[0.14]{m/s}$ at $P_S = \unit[-95.8]{dBm}$).  
Furthermore, Fig.~\ref{fig:PS_over_tag-speed_various-Mod-and-BLF} illustrates performance across data rates between these two extremes.

An alternative approach to increase the signal duration is switching from the 96-bit EPC code to the 128-bit or 256-bit EPC code, which is supported by most RFID tags.  
Fig.~\ref{fig:PS_over_tag-speed_Impinj_lengths} depicts this effect.  
The extended signal duration of longer codes significantly improves accuracy for a given BLF and modulation scheme, but at the cost of a lower capacity.

\section{Conclusions\protect\label{sec:Conclusions}}

In this paper, we analyzed the theoretical limits of Doppler-based motion detection in UHF-RFID systems. 
After deriving the required estimation precision for distinguishing between static and moving tags in Section \ref{sec:Minimum_Required_Estimation_Precision}, we established a theoretical limit on the achievable precision of Doppler shift measurements for RFID tags in Section \ref{sec:MCRBSection}. 
This limit, based on the Modified Cramér-Rao bound~\cite{MCRBAndrea}, is expressed in closed form and applies to both single and dual-segment tag signal measurements. Simulations confirm the bound's tightness.

These results provide insights into the expected estimation performance of Doppler shift measurements in UHF-RFID systems. 
The bounds illustrate how factors such as signal duration, tag signal strength, reader noise figure, and pause length (for dual-segment signals) influence performance.

Importantly, these bounds enable the optimization of systems where Doppler shift is measured, particularly by refining the communication setup between reader and tags. 
For example, they predict how estimation precision improves with increasing tag signal duration and pause length in dual-segment measurements.

In Section \ref{sec:Boundaries-of-the-Motion-Detection}, we derived a theoretical bound for differentiating static and moving tags, presented as a closed-form expression dependent on parameters such as tag signal strength, received noise, signal timing, noise figure, expected tag speed, and required differentiation reliability. We provide results for realistic UHF-RFID signal timings and discuss both theoretical limits under optimal conditions and practical constraints, considering the RFID reader's receiver chain.

This work opens several avenues for future research.
First, further studies are needed to evaluate the impact of multipath propagation on the Doppler shift estimation.
Additionally, developing algorithms to suppress the carrier signal and achieve maximum Doppler shift estimation precision would help assess how closely theoretical bounds can be reached in real-world RFID readers.
Furthermore, geometric effects should be investigated, as real-world tag movement is rarely perfectly aligned with the reader antenna. Ultimately, this work serves as a foundation for optimizing reader-tag communication, including determining the optimal pause length between tag reads based on the derived bounds.

\appendices{}

\section{Estimation of UHF-RFID Reader Noise Figure\label{Sec:NoiseFigureEst}}
In order to have a reference for the experienced noise in UHF-RFID readers, we do an assessment of that of the state-of-the-art RFID reader Impinj R700\footnote{Impinj R700 Series RAIN RFID Readers Datasheet, 2023, Version 2.1, 
available: https://support.impinj.com/hc/article\_attachments/24116880376723 (accessed: March 12, 2025)}.

For optimal detection for binary orthogonal signaling, the lower
bound of the achievable bit-error-rate (BER) for decoding a signal
with a certain $E_{b}/N_{0}$ is given by \cite[(4.2-42)]{Proakis2007}
\begin{equation}
BER=Q\left(\sqrt{\dfrac{E_{b}}{N_{0}}}\right)=\dfrac{1}{2}-\dfrac{1}{2}\textrm{\ensuremath{\erf}}\left(\dfrac{\sqrt{\dfrac{E_{b}}{N_{0}}}}{\sqrt{2}}\right).\label{eq:BER}
\end{equation}
In (\ref{eq:BER}), $\erf(x)$ is the error function, and in the following, $\erf^{-1}(x)$ will denote the inverse error function.

The $P_{S}/N_{0}$ can be calculated from $E_{b}/N_{0}$ using $E_{b}=P_{S}T_{b}$.
Here, $T_{b}$ is the length of one bit. In the case of UHF-RFID data encodings according to \cite{Spec}, it is given by $T_{b}=M/BLF$.
Using this in (\ref{eq:BER}) yields
\begin{equation}
\dfrac{P_S}{N_0} =   \dfrac{2 BLF \left(\textrm{\ensuremath{\erf^{-1}}}\left(1-2\times BER\right)\right)^{2}}{M} . \label{eq:N0-from-PER}
\end{equation}
This term depends on two parameters of the UHF-RFID standard that
can be configured by the reader: The Backscatter Link Frequency, $BLF$,
and the encoding, determined by $M$. 

We now assess $N_0$ for the Impinj R700.  
The corresponding "Reader Modes" document\footnote{Impinj R700 Series Reader Modes, 2024, Version 2.0. Available: \url{https://support.impinj.com/hc/article_attachments/36496556756499} (accessed: March 12, 2025)} lists all supported modes along with the minimum required $P_S$ to achieve a bit error rate (BER) below $10^{-3}$.  
For the most sensitive configuration in the European $\unit[868]{MHz}$ band -- Mode 290 ($BLF = \unit[160]{kHz}$, Miller-8) -- the required received tag signal power is $P_S = \unit[-95.8]{dBm}$.  
Using the parameters $BLF = \unit[160]{kHz}$, $M = 8$, $P_S = \unit[-95.8]{dBm}$, and $BER = 10^{-3}$ in (\ref{eq:N0-from-PER}), and solving for $N_0$ yields $N_0 = \unit[-148.6]{dBm\text{-}Hz}$.  
This value will serve as a reference for a realistic noise spectral density throughout this work.

It should be noted that this value does not represent the actual noise spectral density of the reader.  
Degradations of the received tag signal -- caused by factors such as imperfect carrier suppression, interference, and other impairments -- are effectively included in $N_0$.  
Nevertheless, assuming $N_0 = \unit[-148.6]{dBm\text{-}Hz}$ proves useful for assessing the achievable motion detection performance of RFID readers.

\section{Proof of Theorem \ref{thm:sigma_max}\protect\label{sec:Appendix-required_estimation_precision}}

In case of a normally distributed estimation error, the probability
of a false detection is given by \cite{Proakis2007}
\begin{equation}
P_{err}=0.5\left(1+\textrm{\ensuremath{\erf}}\left(\frac{x-\mu}{\sqrt{2\sigma^{2}}}\right)\right).\label{eq:P_err}
\end{equation}
This is identical for falsely detecting a moving tag as stationary,
and for falsely detecting a stationary tag as moving. In (\ref{eq:P_err}),
$\textrm{\ensuremath{\erf}}(x)$ denotes the error function, $x$
is the decision threshold, $\mu$ the mean value, and $\sigma^{2}$
the variance of the considered random variable. Therefore, $P_{err}$
is the result of integrating the probability density function of a
normally distributed random variable with mean value $\mu$ and variance
$\sigma^{2}$ from $-\infty$ to $x$. In our case, the mean of the
random variable is given by 
\begin{equation}
\mu=f_{D},\label{eq:mu}
\end{equation}
where $f_{D}$ denotes the Doppler shift of objects that are moving
with speed $v$. The decision threshold is given by 
\begin{equation}
x=\dfrac{f_{D}}{2},\label{eq:x}
\end{equation}
which is half of the expected Doppler shift of moving tags. We are
interested in the resulting estimation variance for a fixed $P_{err}$.
Solving (\ref{eq:P_err}) for $\sigma^{2}$ and using (\ref{eq:mu}),
(\ref{eq:x}) and (\ref{eq:doppler_shift_formula}) results in (\ref{eq:required_estimation_variance-general}).
\qedsymbol

\section{Proof of Theorem \ref{thm:MCRB1}\protect\label{sec:Appendix-single_tag_response}}

The proof of Theorem \ref{thm:MCRB1} is split into two parts. Section \ref{sec:proof_MCRB1} provides the derivation of the bound for one signal part and Section \ref{sec:proof_MCRB2} extends it for two signal parts. 

\subsection{One Signal Part: Proof of (\ref{eq:MCRB1_with_modulation_loss_simplified_new}) with (\ref{eq:CT1})\label{sec:proof_MCRB1}} 

The MCRB for a signal of length $T_{0}$ is given by \cite[(25)]{MCRBAndrea},
which, using $P_{S}=\frac{E_{S}}{T_{S}}$ and $T_{0}=LT_{S}$, yields
\begin{equation}
MCRB(\nu)=\dfrac{3}{2\pi^{2}T_{0}^{3}}\frac{N_{0}}{P_{S}}.\label{eq:MCBR1_without_modulation_loss}
\end{equation}

The derivation of (\ref{eq:MCBR1_without_modulation_loss}) assumes that the tag signal is a continuous
signal with constant power, which is only the case for PSK modulation. In the case of ASK, the tag switches between two states, where the received tag signal power is $\sqrt{2P_S}$ in "one state" and zero in "zero state". However, we prove that (\ref{eq:MCBR1_without_modulation_loss}) is also valid in our case for ASK in the following:

The FM0 and Miller encoding schemes have the property that the tag reflects the CW signal quite precisely half of the time.
To consider this, we slightly modify
the derivations in \cite[Sec. III-B]{MCRBAndrea}. 
For this purpose, we replace equation \cite[(12)]{MCRBAndrea} with an ASK RFID signal model.
In this model, the RFID signal consists of a sequence of $L$ symbols
of length $T$. 
Each symbol is in the reflect state with amplitude $\sqrt{2P_S}$ for the first half of the symbol, and in the absorb state with amplitude 0 for the remaining symbol time. 
Mathematically, this leads to:
\begin{align}
s(t)= &  \sqrt{2P_S}\exp\Bigl(j2\pi\nu(t-t_{0})\Bigr)\label{eq:rfid_signal_st}\\
 & \times\sum_{l=0}^{L-1}\rect\Bigl(2\left(t-t_{0}-l\cdot T-T/2\right)/T\Bigr) +n(t)\nonumber,
\end{align}
where $\rect(x)$ is the rectangular function given by
\[
\rect(x)=\begin{cases}
1 & \left|x\right|<1/2\\
0 & \textrm{otherwise}
\end{cases}.
\]
Furthermore, $\nu$ describes the desired Doppler shift, while $t_{0}$ is a random time offset. The modified equation \cite[(17)]{MCRBAndrea}
then leads to

\begin{align}
\int\limits_{T_{0}} & \left|\frac{\partial s(t)}{\partial\nu}\right|^{2}dt= 8 \pi^{2}P_S\sum_{l=0}^{L-1}\left[\int_{l\cdot T}^{l\cdot T+T/2}(t-t_{0})^{2}dt\right]\nonumber \\
 & = 8 \pi^{2}P_S\sum_{l=0}^{L-1}\left[\frac{(t-t_{0})^{3}}{3}\right]_{l\cdot T}^{l\cdot T+T/2}\nonumber \\
 & = 8 \pi^{2}P_S\sum_{l=0}^{L-1} \left( \frac{1}{2}(lT-t_0)^2T + \frac{1}{4}(lT-t_0) T^2 + \frac{T^3}{24} \right) 
\end{align}
The sum can be rearranged and split up into three parts that can be calculated separately using by the summation identities for natural numbers, quadratic numbers and cubic numbers, which yields
\begin{align}
\int\limits_{T_{0}} \left|\frac{\partial s(t)}{\partial\nu}\right|^{2}dt= & \pi^2 P_S   \frac{LT}{3} \Bigl(4L^2T^2 - 3LT^2 - 12LTt_0 \nonumber \\
 &  + 6Tt_0 + 12t_0^2 \Bigr).
\label{eq:MCRB-derivation-modulation}
\end{align}

In order to find the smallest possible variance, the next step is
to find the $t_{0}$ for which (\ref{eq:MCRB-derivation-modulation})
is minimized. This is done by searching for the zero of the first-order
derivative with respect to $t_{0}$. This yields 
\begin{equation}
\frac{d\int_{T_{0}}\left|\dfrac{\partial s(t)}{\partial\nu}\right|^{2}dt}{dt_{0}}= \pi^{2}P_S\left(2LT^2-4L^2T^2+8LTt_0\right),\label{eq:int_t0_precise-1-1}
\end{equation}
which is zero for

\begin{equation}
t_{0}=\frac{T}{2}\left(L-\frac{1}{2}\right).\label{eq:t_0_MCRB1}
\end{equation}
Using (\ref{eq:t_0_MCRB1}) in (\ref{eq:MCRB-derivation-modulation})
yields

\begin{equation}
\int_{T_{0}}\left|\frac{\partial s(t)}{\partial\nu}\right|^{2}dt= \frac{P_S\pi^{2}L^{3}T^{3}}{3}\left(1-\frac{3}{4L^{2}}\right).\label{eq:int_t0_precise-1-1-1-1}
\end{equation}
Using $L^{3}T^{3}=T_{0}^{3}$ and integrating
(\ref{eq:int_t0_precise-1-1-1-1}) into a corrected version of equationn \cite[(16)]{MCRBAndrea}
($N_{0}/2$ is correct instead of $N_{0}$), we finally obtain: 

\begin{equation}
MCRB_{ASK}(\nu)=\dfrac{3}{2\pi^{2}T_{0}^{3}}\frac{N_{0}}{P_{S}}\left(1-\frac{3}{4L^{2}}\right)^{-1}\label{eq:MCRB1_with_modulation_loss_extended}
\end{equation}
The term $(1-\frac{3}{4L^{2}})$ converges to one for realistically large $L$. For the shortest considerable tag signal (an RN16 message without pilot tone with $L=23$), it is $0.9986$.
Therefore, it is justified to say that (\ref{eq:MCRB1_with_modulation_loss_extended}) converges to
\begin{equation}
MCRB_{ASK}(\nu)=\dfrac{3}{2\pi^{2}T_{0}^{3}}\frac{N_{0}}{P_{S}}\label{eq:modulation_loss}
\end{equation}
for realistic $L$, which is the same as the bound for PSK given by (\ref{eq:MCBR1_without_modulation_loss}).

Therefore, the bound for the estimation variance of the Doppler shift estimation based on one tag signal part
is for both PSK and ASK given by (\ref{eq:MCRB1_with_modulation_loss_simplified_new}) with (\ref{eq:CT1}).
\qedsymbol

\subsection{Two Signal Parts: Proof of (\ref{eq:MCRB1_with_modulation_loss_simplified_new}) with (\ref{eq:CT2})\label{sec:proof_MCRB2}} 

In order to reduce the length of the equations, $\Delta T=T_{1}+T_{pause}$
is used in the following calculation. Then, we can calculate:
\begin{align}
\int\limits_{T_{0}} & \left|\frac{\partial s(t)}{\partial\nu}\right|^{2}dt=\nonumber \\
= & 8\pi^{2}P_S\Biggl(\int\limits_{0}^{T_{1}}\left(t-t_{0}\right)^{2}dt+\int\limits_{\Delta T}^{\Delta T+T_{2}}\left(t-t_{0}\right)^{2}dt\Biggr)\nonumber \\
= & 8\pi^{2}P_S\biggl(t_{0}^{2}\left(T_{1}+T_{2}\right)-t_{0}\left(T_{1}^{2}+T_{2}^{2}+2T_{2}\Delta T\right)\nonumber \\
 & \quad+\dfrac{T_{1}^{3}}{3}+\dfrac{T_{2}^{3}}{3}+T_{2}^{2}\Delta T+T_{2}\Delta T^{2}\biggr).\label{eq:MCRB2-integration-1}
\end{align}
To find the minimal possible variance, (\ref{eq:MCRB2-integration-1})
has to be minimized with respect to $t_{0}$. Therefore, we have to
search for the zero of the first-order derivative with respect to
$t_{0}$. This yields

\begin{eqnarray}
\dfrac{d\int\limits_{T_{0}}\left|\dfrac{\partial s(t)}{\partial\nu}\right|^{2}dt}{dt_{0}} & = & 8\pi^{2}P_S\Bigl(2t_{0}\left(T_{1}+T_{2}\right)-T_{1}^{2}\nonumber \\
 &  & \qquad-T_{2}^{2}-2T_{2}\Delta T\Bigr),
\end{eqnarray}
which is zero for 
\begin{equation}
t_{0}=\dfrac{T_{1}^{2}+T_{2}^{2}+2T_{2}\Delta T}{2(T_{1}+T_{2})}.\label{eq:MCRB2-t_0}
\end{equation}
Using (\ref{eq:MCRB2-t_0}) in (\ref{eq:MCRB2-integration-1}) yields

\begin{multline}
\int\limits_{T_{0}}\left|\dfrac{\partial s(t)}{\partial\nu}\right|^{2}dt=\\
=8\pi^{2}P_S\Biggl(\frac{\left(T_{1}+T_{2}\right)^{^{3}}}{12}-\frac{T_{1}T_{2}\left(T_{1}-\Delta T\right)\left(T_{2}+\Delta T\right)}{T_{1}+T_{2}}\Biggr).\label{eq:MCRB2-t_0-1}
\end{multline}
Taking into account $\Delta T=T_{1}+T_{pause}$,
and integrating (\ref{eq:int_t0_precise-1-1-1-1}) into the corrected
equation \cite[(16)]{MCRBAndrea} ($N_{0}/2$ is correct instead of $N_{0}$),
we obtain (\ref{eq:MCRB1_with_modulation_loss_simplified_new}) with $C_T$ given by (\ref{eq:CT2}).

Eq. (\ref{eq:MCRB1_with_modulation_loss_simplified_new}) with $C_T$ given by (\ref{eq:CT2}) can be verified considering
that it simplifies to equation \cite[(22)]{HansMartinPaper} for $T_{1}=T_{2}=T_{N}$,
and to (\ref{eq:MCBR1_without_modulation_loss})
for $T_{pause}=0$ and $T_{0}=T_{1}+T_{2}$.

In the previous section, we have shown that (\ref{eq:MCRB1_with_modulation_loss_simplified_new}) is valid for both PSK and ASK modulation.
We can consider the modulation of the tag signal in the same way we did in the case of one tag signal in the previous subsection.
The number of symbols per signal part in RFID systems is quite large,
also in the case of two distinct signal parts. Therefore, it is a
valid assumption that $MCRB_{ASK}=MCRB$ holds here in the same way as shown by (\ref{eq:modulation_loss}) for the case of one signal part.
Then, the bound for the estimation variance of the Doppler shift estimation based on two tag signal parts
is given by (\ref{eq:MCRB1_with_modulation_loss_simplified_new}) with (\ref{eq:CT2}).
\qedsymbol

\section*{Acknowlegment}

\bibliographystyle{IEEEtran}
\bibliography{IEEEabrv,IEEEexample,literatur}

\begin{IEEEbiography}[{\includegraphics[width=1in,height=1.25in]{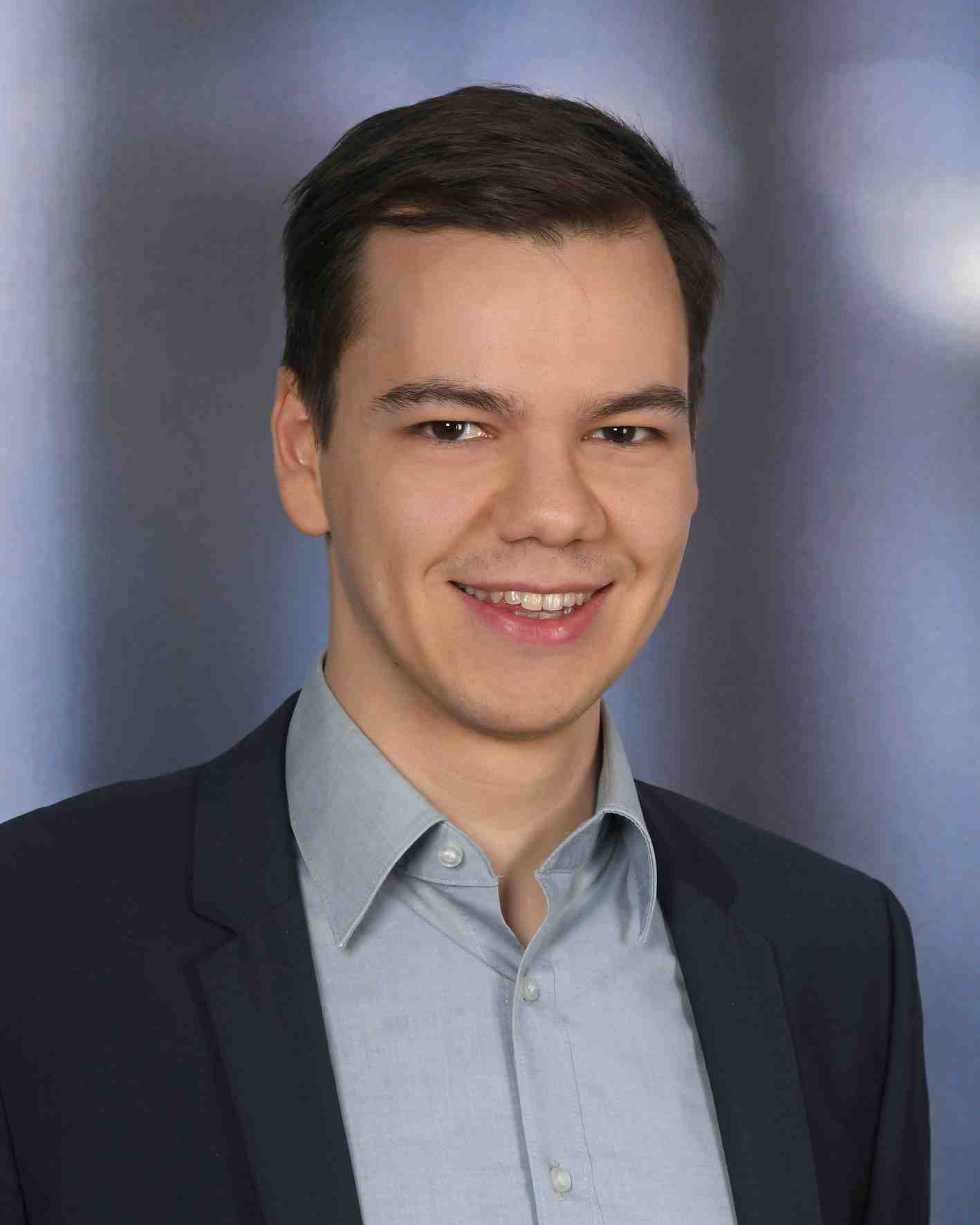}}]{Clemens Korn}
 received the B.Sc. and M.Sc. degrees in Electrical Engineering from
Friedrich-Alexander University Erlangen-Nuremberg in 2015 and 2017,
respectively. In 2017 he joined Fraunhofer IIS, where he conducted
research and various industry projects with focus on UHF-RFID and
other IoT Systems. Further, he was involved in 3GPP standardization
for Fraunhofer IIS as RAN1 delegate for the RedCap study item. In
2021, he started a doctorate at TU Ilmenau, where he conducts research
on Zero-Energy Communications, while continuing a part-time affiliation with Fraunhofer IIS.
\end{IEEEbiography}

\begin{IEEEbiography}[{\includegraphics[width=1in,height=1.25in]{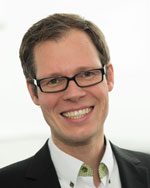}}]{Joerg Robert}
 studied Electrical Engineering and Information Technology at TU
Ilmenau and TU Braunschweig, Germany. From 2006 to 2012, he conducted
research on digital television at the Institute of Communications
Engineering at TU Braunschweig, where he played a key role in the
development of DVB-T2, the second generation of digital terrestrial
television. In 2013, he completed his PhD at TU Braunschweig on the
topic of Terrestrial TV Broadcast using Multi-Antenna Systems.

In 2012, he joined the LIKE Chair at Friedrich-Alexander-Universität
Erlangen-Nürnberg, Germany, where he pursued research on Low Power
Wide Area Networks (LPWAN). From 2021 to 2024, he served as a full
professor and head of the Group for Dependable Machine-to-Machine
Communication at the Department of Electrical Engineering and Information
Technology at Technische Universität Ilmenau, Germany. Since 2025,
he is a full professor specializing in localization systems at the
LIKE Chair at Friedrich-Alexander-Universität Erlangen-Nürnberg. Additionally,
he is affiliated with the Fraunhofer Institute for Integrated Circuits
(IIS) in Nürnberg.

Joerg Robert is actively involved in international standardization
efforts. He currently serves as the secretary of the IEEE 802.15 standardization
group and vice-chair of several wireless standardization groups within
IEEE 802.
\end{IEEEbiography}
\enlargethispage{-1.5cm}

\end{document}